\documentclass[aps,superscriptaddress,showpacs,nofootinbib,eqsecnum,prd,notitlepage]{revtex4} 
\pdfoutput=1
 
\usepackage{epsf}  
\usepackage{eucal} 
\usepackage{amssymb}
\usepackage{amsmath} 
\usepackage{graphicx} 
\usepackage{bm}
\usepackage{dcolumn} 
\usepackage{epsfig} 
\usepackage{multirow}
\usepackage{hyperref}
\usepackage{color}
\usepackage{float}
\usepackage{bm} 
\usepackage[latin1]{inputenc}
\usepackage{ifpdf}
\usepackage{pgf}
\usepackage{xypic}
\usepackage{epstopdf}
 \usepackage{latexsym}

\begin{document}

\title{Matter creation and cosmic acceleration}

\author{Rudnei O. Ramos} \email{rudnei@uerj.br} \affiliation{Departamento de
  F\'{\i}sica Te\'orica, Universidade do Estado do Rio de Janeiro, 20550-013
  Rio de Janeiro, RJ, Brazil}
  
\author{Marcelo Vargas dos Santos}
\email{vargas@if.ufrj.br}
\affiliation{Instituto de F\'{\i}sica, Universidade Federal do Rio de Janeiro, C.P. 68528, 
21941-972 Rio de Janeiro, RJ, Brazil}

\author{Ioav Waga}
\email{ioav@if.ufrj.br}
\affiliation{Instituto de F\'{\i}sica, Universidade Federal do Rio de Janeiro, C.P. 68528, 
21941-972 Rio de Janeiro, RJ, Brazil}

\begin{abstract}

We investigate the creation of cold dark matter (CCDM) cosmology as 
an alternative to
explain the cosmic acceleration. Particular attention is given to the
evolution of density perturbations and constraints coming from recent
observations.  By assuming negligible effective sound speed we compare 
CCDM predictions with 
redshift-space-distortion based $f(z)\sigma_8(z)$ measurements.
We identify a subtle issue associated with which contribution in the 
density contrast should be used
in this test and then show that the CCDM results are the same as those
obtained with $\Lambda$CDM. These results
are then contrasted with the ones obtained at the background level. 
{}For the background tests
we have used type Ia supernovae data (Union 2.1  compilation)
in combination with baryonic acoustic oscillations and cosmic
microwave background observations and also measurements of the
Hubble parameter at different redshifts. As a consequence of the studies we have
performed at both the background and perturbation levels, we explicitly show that
CCDM is observationally degenerate with respect to $\Lambda$CDM (dark degeneracy). 
The need to overcome the lack of a fundamental microscopic basis for the CCDM is the 
major challenge for this kind of model.

\end{abstract}


\maketitle

\section{Introduction}

Explaining the recent cosmic acceleration, believed to be related
to some form of dark energy, and supported by the observations of high
redshift supernovae and by other independent observational data, such
as the results coming from cosmic microwave background radiation
(CMBR) and  with baryonic acoustic oscillation (BAO), is one of the
present day  challenges in cosmology.  Among the many possible
proposals (for recent reviews, see, e.g., Ref.~\cite{DEreviews} and
also references therein), possibly a cosmological constant ($\Lambda$)
is the simplest answer to explain the late-time cosmic
acceleration. Of course, this also brings  some theoretical
difficulties, which are how to explain its origin, right magnitude and
why it comes to dominate just now.

Just like the early-time cosmic acceleration associated with
inflation,  a negative pressure can be seen as a possible driving
mechanism for the  late-time accelerated expansion of the Universe as
well.  One of the earliest alternatives that could provide a mechanism
producing such accelerating phase of the Universe is through a
negative pressure produced by viscous or particle production
effects. {}For instance, one of the first works relating particle
production, in particular as a result of a nonstationary
gravitational field and that can be described phenomenologically by
means of a negative pressure, is due to
Zeldovich~\cite{Zeldovich:1970si}. This is much similar to the idea
put forward by Murphy~\cite{Murphy:1973zz} and also later by
Hu~\cite{hu1982}, that particle production might also be described
equivalently in terms of a bulk viscous pressure in the cosmological
fluid.  In this context, since a bulk viscous pressure is a negative
pressure contribution in the  energy-momentum stress
tensor~\cite{weinberg}, it has lead to an extensive literature on
applications related to bulk viscous cosmology  (for a partial sample
of the earliest works on bulk viscous cosmologies, see for example
Refs.~\cite{Murphy:1973zz,hu1982,othersbulk, lima88}). In addition,
more recently, there has also been a surge of interest in exploring
the effects of the bulk pressure as the origin of the present
accelerated expansion of  the Universe (see, e.g.,
Ref.~\cite{bulkDE}).
A closed related scenario to the bulk viscous cosmology is that of the
so-called adiabatic matter creation, which makes use of ideas of the
thermodynamics of open systems in the context of cosmology and
initiated by Prigogine and collaborators~\cite{prigogine}.   A
covariant formalism approach has later been formulated in Ref.~\cite{calvao}. 

Despite the fact that bulk viscous and matter creation cosmologies
apparently look similar,  they have some fundamental differences.
Bulk viscous cosmologies are associated with a generalization of  the
hydrodynamics of ideal fluids for the case of nonideal ones, with
constitutive equations describing the viscous pressures built as
additional correction terms to the equilibrium energy-momentum  stress
tensor~\cite{weinberg}. As such, the viscous pressure contributions
can be  seen as small nonequilibrium contributions for the
energy-momentum tensor for nonideal fluids. It happens, however, that
most of the effects of a bulk viscous pressure to cosmology, as for
example when it is used as a mechanism for inflation, it typically
requires an extrapolation beyond the limit of validity for these
theories~\cite{pacher} (see, however, Ref.~\cite{lima88}). In the
context of matter creation, even though  also a negative effective
pressure can be associated with it, there is in principle no such
limitation as with a negative bulk viscous pressure.  

Particle creation models~\cite{prigogine,calvao,mattercreation,freaza,zimdahl,lima1,basilakos,jesus,lima2}, as
the one treated in this work, should also
not be confused with other cosmological scenarios where particle
production is present, like, e.g., in warm inflation~\cite{BMR}. In
warm inflation models the inflationary evolution can be strongly
influenced by relativistic (radiation) particle production. In these
models there can also be negative pressure effects as a result of a
bulk viscous pressure from the radiation bath, but these effects are in
general small in the inflationary context~\cite{warmbulk}. 

Many authors have explored scenarios of matter creation in cosmology, but 
here we are particularly interested in the gravitationally 
induced particle creation scenario denominated ``creation of cold dark matter'' (CCDM) 
\cite{lima1,basilakos,jesus,lima2} in which a special
choice of the particle production rate produces a cosmology that, at the
background level, is indistinguishable from
the standard $\Lambda$ cold dark matter ($\Lambda$CDM) model. 
However, as we are going to discuss in this paper, at
the  perturbative level this degeneracy is more subtle and care should be taken when contrasting CCDM
predictions with those obtained in the standard $\Lambda$CDM cosmology.
{}Furthermore, perturbations in the case of CCDM
cosmology have mostly been studied in the context of the so-called
neo-Newtonian formalism~\cite{lima3}. {}Following Ref.~\cite{reis}, here we will
show that this formalism for studying density perturbations has limitations and a fully
relativistic one (like that, for example,
of Ref.~\cite{kodama}) is required in the case in which the effective speed of sound cannot be neglected. 

The rest of this paper is organized as follows. In
Sec.~\ref{thermo} we briefly  review the thermodynamics for matter
creation cosmology. In Sec.~\ref{cosmomodels} we discuss the
background equations and their solutions.  In Sec.~\ref{evol} both
the neo-Newtonian and relativistic approaches are discussed and the
differences between the two are given.  In Sec.~\ref{ccdm} we
analyze the observational constraints on the CCDM model
we have considered here.  {}Finally, our conclusions and final remarks
are given in Sec.~\ref{conclusions}.


\section{Thermodynamics of matter creation in a simple fluid} 
\label{thermo}

Let us briefly review here the thermodynamics of matter creation.
{}For simplicity, we will restrict to the case of a single fluid, but
it can easily be generalized to multiple coupled fluids as well. To
describe the thermodynamic states of a relativistic simple fluid we
use the following macroscopic variables: the energy-momentum tensor
$T^{\alpha \beta }$; the particle flux vector  $N^{\alpha }$; and the
entropy flux vector $s^{\alpha }$.  The energy-momentum tensor
satisfies the conservation law, $T^{\alpha \beta }{}_{;\beta }=0$, and
here we consider situations in which it has the perfect-fluid form

\begin{equation}
T^{\alpha \beta }=\left( \rho +P\right) \,u^{\alpha }
u^{\beta}-P\,g^{\alpha \beta }\;.
\end{equation}
In the above equation $\rho $ is the energy density, $P$ is the
isotropic dynamical pressure, $g^{\alpha \beta }$ is the metric tensor
and  $u^{\alpha}$ is the fluid four-velocity (with normalization
$u^\alpha u_\alpha=1$). 

The dynamical pressure $P$ is decomposed as

\begin{equation}
P=p+\Pi\, ,
\end{equation}
where $p$ is the equilibrium (thermostatic) pressure and $\Pi $ is a
term present  in scalar dissipative processes. Usually, it is
associated with the so-called bulk pressure~\cite{weinberg}. In
the cosmological context, besides this meaning, $\Pi $ can also be
relevant when particle number is not conserved~\cite{prigogine}. In
this case, $\Pi \equiv p_c$ is called the  ``creation pressure''. It
is important to mention that, the bulk pressure, as already mentioned
in the Introduction, can be seen as a correction to the thermostatic
pressure when near to equilibrium, thus, it should be always smaller
than the thermostatic pressure, $\left| \Pi \right| < p$. This
restriction, however, does not apply for the creation pressure. So,
when we have matter creation, the total pressure $P$ may become
negative and, in principle, drive an accelerated expansion.

The particle flux vector is assumed to have the following form

\begin{equation}
N^{\alpha }=n\,u^{\alpha }\,,
\end{equation}
where $n$ is the particle number density. $N^{\alpha }$ satisfies the
balance equation $N^{\alpha }\,_{;\,\alpha }= n \Gamma $,  where
$\Gamma$ is the particle production rate. If $\Gamma >0$, we have
particle creation,  particle destruction occurs when $\Gamma <0$ and
if $\Gamma =0$ particle number is conserved.

The entropy flux vector is given by

\begin{equation}
s^{\alpha }=n\,\sigma u^{\alpha }\,,
\end{equation}
where $\sigma$ is the specific (per particle) entropy. Note that the
entropy must satisfy the second law of thermodynamics $s^{\alpha
}\,_{;\,\alpha }\geq 0$.  Here we consider adiabatic matter creation,
that is, we analyze situations in which $\sigma $ is constant. With
this condition, by using the Gibbs relation, it follows that the
creation pressure is related to $\Gamma$ by~\cite{prigogine,calvao},

\begin{equation}
p_c =-\,\frac{\rho +p}{3\,H}\,\Gamma\;,  
\label{pc}
\end{equation}
where $H=\dot{a}/a$ is the Hubble parameter, $a$ is the scale factor
of the {}Friedmann-Robertson-Walker (FRW) metric [see Eq.~(\ref{frw})
below] and the overdot means differentiation with respect to the
cosmic time. It is also straightforward to show that, if $\sigma $ is
constant, the second law of thermodynamics implies that $\Gamma \geq
0$ and, as a consequence, particle destruction ($\Gamma <0$) is
thermodynamically forbidden~\cite{prigogine,calvao}.  Since $\Gamma
\geq 0$, it follows from Eq.~(\ref{pc}) that, in an expanding universe
($H>0$), the creation pressure $p_c$ cannot be positive.


\section{Cosmological models with particle creation}
\label{cosmomodels}

Before we discuss the evolution of linear perturbations in
cosmological models with matter creation, we first consider their
background equations. By assuming spatial homogeneity and isotropy,
which is a good approximation at large scales, we are lead to the FRW
line element,

\begin{equation}
ds^{2}=dt^{2}-a^{2}\left( t\right) \left[ \frac{dr^{2}}{1-kr^{2}}
  +r^{2}\left( d\theta ^{2}+\sin ^{2}\theta \,d\varphi ^{2}\right)
  \right]\, . \label{frw}
\end{equation}
Here  $k=0,\pm 1$ characterizes the curvature of the spatial sections
of space-time and we  are assuming $c=1$, as usual.  {}For the sake of 
simplicity, from now on we
also assume flat space ($k=0$), which is in good agreement with CMBR
observations.  In this paper we are mainly interested in processes that
occurred after radiation domination.  Therefore, as a first
approximation, we neglect radiation and, for the sake of simplicity,
we also neglect baryons  considering only the presence (and creation)
of pressureless ($p=0$) dark matter particles.  

The Einstein equations for the models we consider can be expressed
simply as

\begin{equation}
H^{2}=\left( \frac{\stackrel{.}{a}}{a}\right) ^{2}=\frac{8\pi
  G}{3}\rho\,,   
\label{H2}
\end{equation}

\begin{equation}
\frac{\stackrel{..}{a}}{a}=\,\stackrel{\,.}{H}+\,H^{2}=-\frac{4\pi
  G}{3} \left( \rho +3p_c \right)\,.  
\label{addot}
\end{equation}
To the above equations we add Eq.~(\ref{pc}) (with $p=0$) to get,

\begin{equation}
\stackrel{.}{\rho }+\,3\,H\, \rho  =\rho\, \Gamma\,. 
\label{rho}
\end{equation}

In order to integrate the above equations, it is necessary to assume a
special form for $\Gamma$. Several models that have previously been
studied in the literature can all be generalized by the following
expression for the  particles production rate~\cite{freaza}:

\begin{equation}
\Gamma =3\,\beta \,H_{0}\left( \frac{H}{H_{0}}\right) ^{\alpha }\,,
\label{Gamma}
\end{equation}
where $\alpha $ and $\beta $ are $\mathcal{O}(1)$ dimensionless
constants and $H_{0}$ is the present value of the Hubble
parameter. Throughout this paper we use the subscript ``$0$'' to
denote the present value of quantities.  {}From the above equations,
we get the following differential equation for $H$:

\begin{equation}
\frac{dH}{dz}\,(1+z)=\frac{3}{2}\,H_{0}\left[ \frac{H}{H_{0}}-\beta
  \left(  \frac{H}{H_{0}}\right)^{\alpha }\right]\,,  
\label{dHdt}
\end{equation}
where $z=1/a-1$ is the redshift. The above equation can easily be
integrated, leading to the result~\cite{freaza}

\begin{eqnarray}
H=\begin{cases}H_{0}\left[ \beta +\left( 1-\beta \right) \left(
  1+z\right)^{\frac{3}{2} \left( 1-\alpha \right)
  }\right]^{\frac{1}{1-\alpha }} \,, & \text{if} \,\,\, \alpha \neq
1\, ,\\  H_{0}  \left(1+z\right)^{\frac{3}{2}\left( 1-\beta \right) }
\,, & \text{if} \,\,\, \alpha =1\;. 
\end{cases}
\label{Half}
\end{eqnarray}

{}From now on we focus on  the particular case  $\alpha=-1$ in
Eq.~(\ref{Gamma}). {}Following Ref.~\cite{lima2}, we refer to this
model as ``creation  of cold dark matter'' (CCDM). With $\alpha=-1$,
from Eq.~(\ref{Half}), we obtain 

\begin{equation}
\frac{H^2}{H_{0}^2}=\left(1-\beta \right)  \left( 1+z\right)
^3+\beta\,.
\label{Lcdm}
\end{equation}
The above equation indicates that the expansion rate $H$ in CCDM has
the same exact form as in  flat $\Lambda$CDM models with $\beta$
playing the role of the cosmological constant density  parameter at
present time~\cite{zimdahl,freaza}, $\Omega_{\Lambda0}$. Notice that,
by using Eq.~(\ref{H2}),  the expression for the particle production
rate Eq.~(\ref{Gamma})  can be written as~\cite{lima1}

\begin{equation}
\Gamma=\frac{3\beta
  H_0^2}{H}=3\beta\left(\frac{\rho_{c0}}{\rho}\right) H\;,
\end{equation}
where $\rho_{c0}\equiv 3 H_0^2/(8\pi G)$ is the critical density at
present time, which, in our  flat-space and simple-fluid
approximation, is equal to the value of dark matter energy density  at
present. Notice also that for $\alpha=-1$, the creation pressure,
$p_c=- \beta \rho_{c0}$,  is constant and by using Eq.~(\ref{H2}), the
dark matter energy density can be written as

\begin{equation}
\rho=\rho_{c0} \left[ \left( 1-\beta \right)  \left( 1+z\right)^3 +
  \beta \right]\,.
  \label{rho2}
\end{equation}

As remarked above, the CCDM model mimics exactly the
$\Lambda$CDM background expansion history,  so we should expect good
accordance of this  model with kinematic cosmological tests like from
supernovae type Ia (SNIa) and BAO,  that essentially depend only on
distances and, thus, does not depend on the perturbation results. Does
this mimicry remains at the perturbation level ?  Answering this
question is somewhat subtle and we will discuss  it in
Sec.~\ref{ccdm}. 

Another point to be stressed here is that, although in CCDM we  have a
kind of unification of the dark sector, it does not solve or alleviate the
so-called cosmological constant problems. {}For instance, the  
old cosmological constant problem is not solved since, like in
quintessence, $\Lambda$ is assumed to be zero from the
beginning. Of course we hope that this problem will be resolved in the
context of quantum field theory and not by cosmology.
{}However, the fine-tuning and the cosmic coincidence problems
are essentially the same as in $\Lambda$CDM. To better understand
this, we now write the total dark matter energy density $\rho$ as 

\begin{equation}
\rho=\rho_{\rm conserved}+\rho_{\rm created},
\end{equation} 
where $\rho_{\rm conserved}=\rho_{c0}  \left( 1-\beta \right)  \left(
1+z\right)^3$ is the conserved part of the dark matter energy density
and $\rho_{\rm created}=\rho_{c0}\beta$ is the created one.  The
cosmological problems  can now be cast as follows: Why was the created
(and constant) part so small (as compared to the energy densities
of other fields) and finely adjusted in the beginning  of the Universe
evolution? Why only at recent times are the conserved and the created (and
constant)  dark matter energy densities comparable ?  Therefore,
CCDM model has essentially the same conceptual
difficulties as $\Lambda$CDM. Indeed, from the theoretical point of view 
the situation is even worse in CCDM.
Although some authors (see, e.g., Refs.~\cite{lima1,jesus,lima2} and
references therein) try to motivate the CCDM
scenario in terms of gravitational particle production in an
expanding universe,  currently there is no fundamental basis 
for the chosen particle production rate and we can only treat CCDM 
as a phenomenological model.
In this context, we adopt a more pragmatic approach and, in the
following sections,  we discuss if observations that depend on the
growth of perturbations can distinguish CCDM from $\Lambda$CDM. If
yes, the CCDM model can be tested.  If it produces results that are
not compatible with the observations, then it can be discarded from
the beginning. If the results are better than the ones produced with
the $\Lambda$CDM, then we can pursue further and look more closely for
the microscopic motivations for the model. However, if the CCDM and
$\Lambda$CDM are observationally degenerated
with each other, then we must resort to the Occam's Razor
principle to guide us. Accordingly, the simplest model  (i.e.,
$\Lambda$CDM) becomes preferable unless further theoretical developments 
change the current situation. 


\section{Evolution of linear density perturbations: neo-Newtonian versus 
relativistic approach} 
\label{evol}

We now turn our attention to the growth of linear perturbation in
matter creation models. {}Following Refs.~\cite{jesus,lima1}, we first
consider it in the neo-Newtonian context.  The idea of a Newtonian
expanding universe was developed by Milne~\cite{milne} and also  by
McCrea and Milne~\cite{mccrea} in the 1930's. By considering a
pressureless fluid and assuming Newtonian dynamics and gravitation, it
was shown that the governing Newtonian differential equations are
identical in form to the relativistic ones. This approach, known as
Newtonian cosmology (NC), is quite helpful in giving insight into the
physical significance of an expanding universe. The NC equations were
generalized to include uniform pressure by McCrea~\cite{mccrea2} in a
paper in which the hypothesis of continuous creation of matter was
investigated. The same equations were reobtained later in
Ref.~\cite{harrison2} in a different context. However, as pointed out
in Ref.~\cite{sachs},  although the Newtonian background evolution
equations with pressure are identical in form to the relativistic ones
(assuming zero spatial curvature), at the perturbative level they are
only equivalent when pressure is zero.  To circumvent this difficulty,
in Ref.~\cite{lima3} it was suggested a modification of the continuity
equation. This formulation, known as  neo-Newtonian approach, has also
limitations, as pointed out in Ref.~\cite{reis}, as we now discuss.

The basic equations that describe the neo-Newtonian formulation
are~\cite{mccrea2,harrison2,lima3}

\begin{eqnarray}
&& \nabla_{r}^{2}\phi   =    4\pi G(\rho+3P)\;,
\label{poisson}
\\ &&\left(\frac{\partial \textbf{u}}{\partial t}\right)_{r} +
(\textbf{u} \cdot \bf{\nabla}_{r})\textbf{u}    = -\bf{\nabla}_{r}\phi
- (\rho + \it{P} )^{-1} \bf{\nabla}_{r} \it{P}\;, 
\label{euler}
\\ &&\left(\frac{\partial\rho}{\partial t}\right)_{r} +
\bf{\nabla}_{r}\cdot(\rho \textbf{u}) + \it{P} \bf{\nabla}_{r}\cdot
\bf{u} = 0\;. 
\label{cont}
\end{eqnarray}
Equations (\ref{poisson}), (\ref{euler}) and (\ref{cont}) are,
respectively, the modified Poisson,  Euler and energy conservation
equations, where relativistic effects of pressure were included. In
the above equations, ${\bf u}$ is the velocity field and $\phi$ is the
gravitational potential of the cosmic fluid.

As usual in perturbation theory~\cite{kodama}, we assume small
perturbations around the homogeneous background solution in the form:
$ \rho = \tilde{\rho} + \delta\rho = \tilde{\rho}(1+\delta)$, $P    =
\tilde{P} + \delta P$, $\phi =  \tilde{\phi} + \varphi$, and
$\textbf{u}  =  H \, \textbf{r} + \textbf{v}$. We use a tilde
,``$\,\,\tilde{}\,\,$'', to denote background quantities.  Introducing
comoving coordinates $\textbf{x}=\textbf{r}/a$, neglecting shear and
vorticity and taking into account the background equations, after some
algebra we can derive the following differential equation for the
density contrast~\cite{reis},

\begin{eqnarray}
&&\ddot{\delta} - \left[ 3\left( 2w-c_s^2 - c_{eff}^2 \right) -2
    \right] H \dot{\delta}  \nonumber \\ &&+  3H^2 \left\{ \left[
    \frac{3}{2} w^2-4w-\frac{1}{2}+3c_s^2\right]+c_{eff}^2  \left(
  3c_s^2-6w-1 \right)+\frac{(c_{eff}^2)^\textbf{.}}{H} +
  \frac{k^2}{a^2} \frac{c_{eff}^2}{3H^2} \right\} \delta
  =0, \label{delta1}
\end{eqnarray}
where $w=\tilde{P}/\tilde{\rho}$, $c_{eff}^2=\delta{P}/\delta\rho$,
$c_s^2=\dot{\tilde{P}}/\dot{\tilde{\rho}}=w-\dot{w}/[3H(1+w)]$ and $k$
is the comoving wave number. We are looking for solutions of the form
$\delta(\textbf{x},t) =\sum\limits_{k} \delta_k(t)e^{i\textbf{k}\cdot
  \textbf{x}}$ and for the sake of simplicity, we have dropped the index
$k$ from $\delta$ in Eq.~(\ref{delta1}). We have also assumed that
$c_{eff}^2$  is a function of time only.

Let us now consider the evolution of density perturbations in a
general-relativistic framework. In this case, following standard
calculations~\cite{kodama},  assuming zero anisotropic pressure
perturbations, besides flat space,   we obtain 

\begin{equation}
\ddot{\Delta} - \left[ 3\left( 2w-c_s^2 \right) -2 \right] H
\dot{\Delta} + 3H^2  \left\{  \left[ \frac{3}{2}
  w^2-4w-\frac{1}{2}+3c_s^2 \right]
+\frac{k^2}{a^2}\frac{c_{s}^2}{3H^2}\right\} \Delta  = -
\frac{k^2}{a^2}w\hat{\Gamma}, \label{Delta}
\end{equation}
where 

\begin{equation}
\hat{\Gamma}\equiv \frac{\delta P}{\tilde{P}}-c_s^2 \frac{\delta
}{w}=\frac{(c_{eff}^2-c_s^2)}{ w} \, \Delta\,,
\end{equation}
is the gauge-invariant entropy perturbation,
$c_{eff}^2\equiv\frac{\delta P}{\delta \rho}|_{rest}$ is the
effective sound speed (defined in the matter rest frame) \cite{hu2}
and the  gauge-invariant quantity $\Delta$ represents the matter
density contrast in the slicing such that the matter four-velocity is
orthogonal to constant time hypersurfaces~\cite{kodama},

\begin{equation}
\Delta= \delta +3 (1+w) H \frac{a}{k} \,(v-B),
\end{equation}
where $v-B$ is associated with the deviation of the matter
four-velocity from the vector normal to the constant time
hypersurfaces.  

To compare the relativistic and neo-Newtonian differential equations
for the density contrast, we go to the rest gauge \cite{hu2}, where
$\Delta=\delta$, and write Eq.~(\ref{Delta}) as

\begin{equation}
\ddot{\delta} - \left[ 3\left( 2w-c_s^2 \right) -2 \right] H
\dot{\delta} + 3H^2  \left\{  \left[ \frac{3}{2}
  w^2-4w-\frac{1}{2}+3c_s^2 \right]
+\frac{k^2}{a^2}\frac{c_{eff}^2}{3H^2}\right\} \delta  =
0. \label{rdelta}
\end{equation}
Therefore, by simple inspection, we see that even for time-independent
$c_{eff}^2$, Eqs.~(\ref{delta1}) and (\ref{rdelta}) are only identical
when the effective sound speed $c_{eff}^2$ is equal to zero.  It
should also be remarked that in the more general case, in which
$c_{eff}^2 \ne 0$, the last term inside the braces in
Eq.~(\ref{rdelta}) can only be neglected in the long-wavelength limit
($k=0$), in which case the Newtonian approximation is not
valid. Therefore, using  Eq.~(\ref{delta1}), assuming $c_{eff}^2 \ne
0$ and neglecting the last term inside the braces is not a correct
procedure (as adopted for example in Ref.~\cite{lima2}), first
because Eq.~(\ref{delta1}) is not valid for $c_{eff}^2 \ne 0$, and
second because the Newtonian approximation is also not valid in the
long-wavelength limit.


\section{The CCDM model: theory versus observations}
\label{ccdm}

\begin{figure}[ht]
\begin{centering}
    \includegraphics[width=0.49\textwidth]{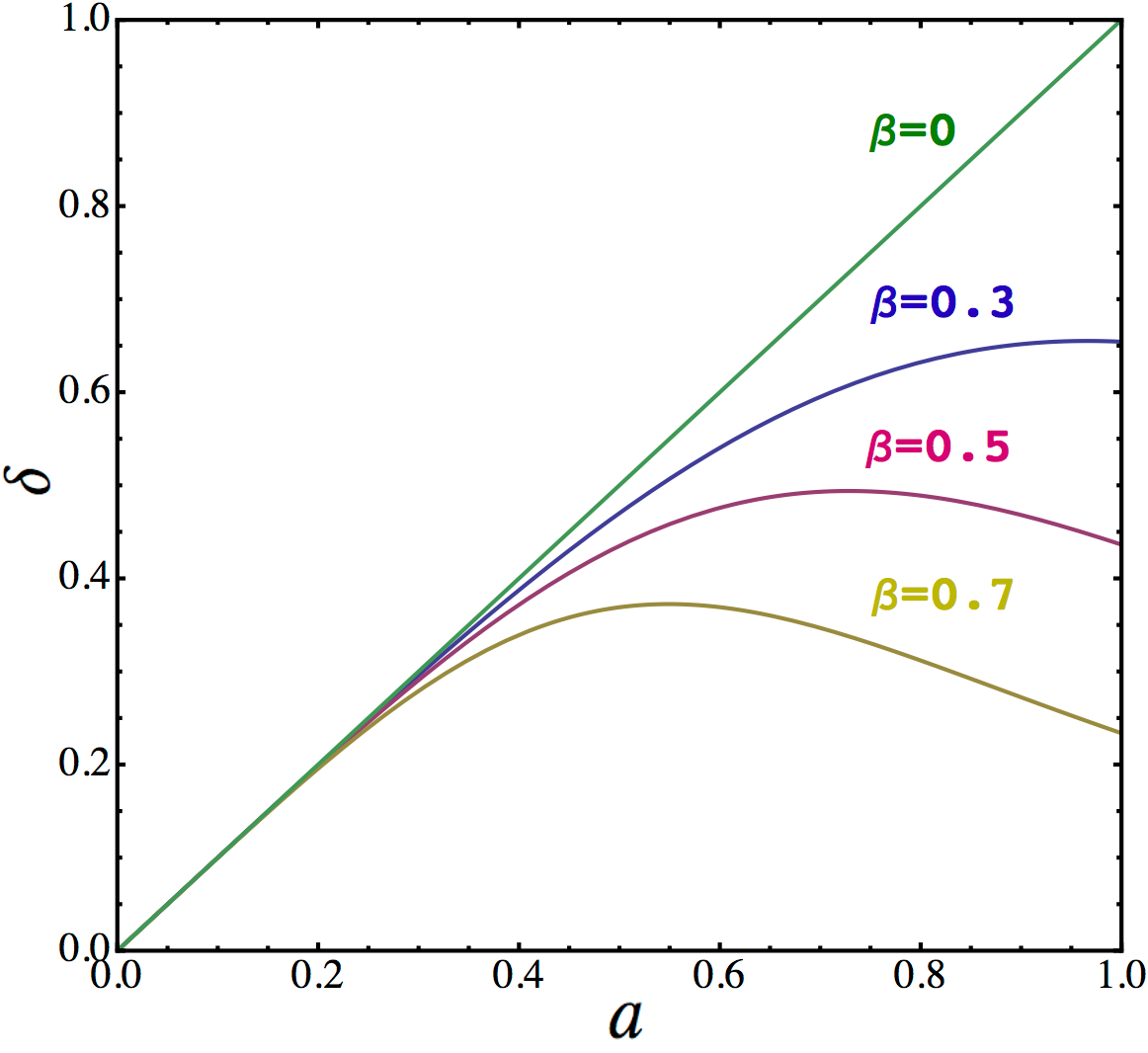}
    \par   \end{centering}
    \caption{The density contrast $\delta$ in CCDM as a function of
      the scale factor $a$, for different values of $\beta$. }
    \label{fig1}
\end{figure}

Let us now consider the observational constraints on the CCDM
model from the linear growth of energy density
perturbation data. {}For this model, the background creation pressure
$\tilde{p}_c = -\beta \rho_{c0}$ is constant and, therefore,
$c_s^2=0$. We first assume $c_{eff}^2=0$, such that the neo-Newtonian
and the general-relativistic approaches are equivalent. Notice that
this corresponds to adiabatic perturbations ($\hat{\Gamma}=0$), since
$c_{eff}^2$ and $c_s^2$ are equal. By changing the variable from the
cosmic time $t$ to the scale factor $a$, recalling Eq.~(\ref{Lcdm})
and that for a constant creation pressure, as we are considering here,
we have $H_0^2 = -w H^2/\beta$, we then obtain that Eq.~(\ref{rdelta})
can be written as 

\begin{equation}
\delta'' + \frac{3}{2a}\left(1-5w \right) \delta' + \frac{3}{2a^2}
\left( 3 w^2-8w -1 \right)\delta = 0\,, 
\label{delta_a}
\end{equation}
where the prime denotes derivative with respect to the scale factor
$a$ and the equation of state parameter is given by
$w(a)=-\beta/\left[\beta+(1-\beta)a^{-3}\right]$. Observe that for
$\beta=0$ ($w=0$), there is no matter creation and the model reduces
to the Einstein--de Sitter model. In the opposite limit, $\beta = 1$
($w=-1$), there is no conserved dark matter and the de Sitter model is
recovered. To integrate Eq.~(\ref{delta_a}), we introduce a new
variable $x=-a^3 \beta/(1-\beta)$ and write the density contrast as
$\delta(x)=a/(1-x) G(x)$. With these definitions we rewrite
Eq.~(\ref{delta_a}) as

\begin{equation}
x(1-x)G''(x) + \left(\frac{11}{6}-\frac{7}{3} x \right)
G'(x)-\frac{1}{3} G(x)=0.
\end{equation} 
The exact solution of the above equation can be expressed in terms of
hypergeometric functions  $_2F_1(a,b;c;x)$ as

\begin{equation}
G(x) = C_1 \; {}_2 F_1 \left(\frac13, 1; \frac{11}{6}; x \right)+ C_2
\;  x^{-\frac56} {}_2F_1\left(-\frac12, \frac16; \frac16;x\right)\,,
\label{hiperg}
\end{equation}
where $C_1$ and $C_2$ are arbitrary constants. The first term on the
right-hand side of Eq.~(\ref{hiperg}), by looking at the asymptotic
behavior of the hypergeometric function, can be identified with the
growing mode, while the second term is a decaying mode.  Neglecting
the decaying mode, we write $\delta$  as

\begin{equation}
\delta(a,\beta) = \frac{a}{1+\frac{a^3\beta}{1-\beta}}\;  {}_2 F_1
\left(\frac13, 1; \frac{11}{6}; -\frac{a^3\beta}{1-\beta}\right),
\end{equation}
where the density contrast is normalized such that for $a\ll 1$ we
have $\delta =a$, since at high redshifts the CCDM behaves like the
Einstein--de Sitter model.  In {}Fig.~\ref{fig1} we show the density
contrast $\delta$ for dark matter in CCDM as a function of the scale factor for
several values of $\beta$. 

\begin{figure}[ht]
\begin{centering}
    \includegraphics[width=0.49\textwidth]{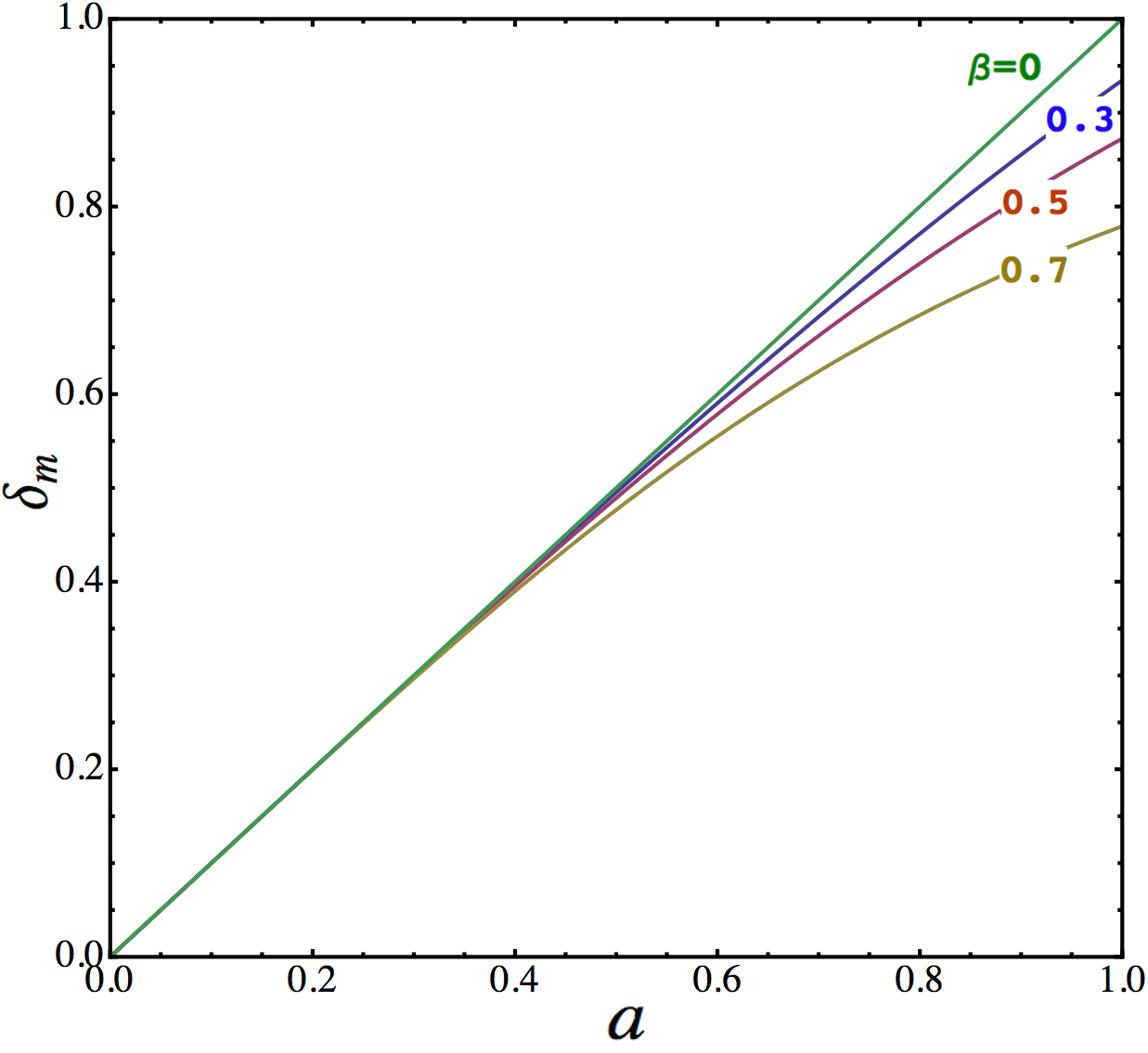}
    \par   \end{centering}
    \caption{The matter density contrast $\delta_m$ in $\Lambda$CDM as
      a function of the scale factor $a$, for different values of
      $\beta=1-\Omega_{m0}$. }
    \label{fig2}
\end{figure}

The density contrast for dark matter ($\delta_m=\frac{\delta
  \rho_m}{\rho_m}$) in a flat $\Lambda$CDM model, such that
$\Omega_{m0}=1-\beta$, can expressed as~\cite{silveira}:

\begin{equation}
\delta_m(a,\beta) = a \; {}_2 F_1\left(\frac13, 1; \frac{11}{6};
-\frac{a^3\beta}{1-\beta}\right), 
\label{deltam}
\end{equation}
In {}Fig.~\ref{fig2} we show the density contrast $\delta_m$ as a
function of the scale factor for several values of $\beta$.  By
comparing  {}Fig.~\ref{fig1} with {}Fig.~\ref{fig2}, it is clear the 
density contrast suppression in CCDM, as we increase $\beta$, when 
compared to the $\Lambda$CDM case. We remark that this suppression 
is stronger than
the one obtained by the authors in Ref.~\cite{roany}, who have used a
different approach, and is in accordance with {}Fig. $1$ of
Ref.~\cite{jesus}.  But we are then left with the question, what is
the origin of this suppression ?  To answer this question note that

\begin{equation}
\frac{\delta}{\delta_m}=\frac{1}{1+\frac{a^3\beta}{1-\beta}}=
\frac{(1-\beta)a^{-3}}{(1-\beta)a^{-3}+\beta}=\frac{\rho_m}{\rho},
\label{deltaratio}
\end{equation}
where $\rho_m =\rho_{c0} \Omega_{m0} a^{-3}$ is the energy density of
dark matter in flat $\Lambda$CDM and $\rho$, given by
Eq.~(\ref{rho2}), is the total CDM energy density in CCDM. Notice that
$\rho_m$ is also equal to $\rho_{cl} =\rho_{c0}(1-\beta)a^{-3}$, the
CDM {\it clustered part} in CCDM. {}From Eq.~(\ref{deltaratio}), we get
that $\delta\rho=\delta \rho_m$ and, therefore, the mentioned
suppression in the density contrast appears, when the constant,  {\it
  nonclustered and created part} of the CCDM energy density starts to
become non-negligible. It is important to keep in mind that matter in
CCDM clusters exactly in the same manner as it does in $\Lambda$CDM,
since the gravitational potential is the same. {}Furthermore, light
also follows the same geodesics and, since we have assumed $c_{eff}^2=0$,
we cannot observationally distinguish CCDM from $\Lambda$CDM. This property is
related to the dark degeneracy~\cite{kunz} and remounts to the
discussion on the $\Lambda$CDM limit of the generalized Chaplygin gas
model~\cite{fabris,avelino}.

The above consideration is particularly relevant when one wants to
compare CCDM model predictions with observations that depend on how
linear perturbations grow. Consider, for instance, the
$f(z)\sigma_8(z)$ test~\cite{song}, where $f(z)$ is the linear growth
rate and $\sigma_8(z)$  is the redshift-dependent root-mean-square
mass fluctuation in spheres with radius $8h^{-1}$ Mpc. In CCDM, which
of the two quantities, $\delta$ or $\delta_m$, should we use in this
test ? Unlike in Ref.~\cite{lima2}, in this work we use
$\delta_m=\delta_{cl}\equiv\frac{\delta \rho}{\rho_{cl}}$ instead of
$\delta=\frac{\delta \rho}{\rho}$. The justification for this choice
is based on the fact that for the $f(z)\sigma_8(z)$ test only {\it
  clustered matter} is important.

To compare CCDM model predictions with observations we use the
redshift-space-distortion based $f(z)\sigma_8(z)$
measurements~\cite{song}, which are displayed in Table~\ref{fsig8}.
The data were obtained by the following surveys:
6dFGRS~\cite{beutler}, 2dFGRS~\cite{percival}, WiggleZ~\cite{blake},
SDSS LRG~\cite{samushia2},   BOSS CMASS~\cite{reid} and
VIPERS~\cite{delatorre}. 

Here $f(z)$ is the linear growth rate given by

\begin{equation}
f(z)\equiv \frac{d\ln \delta_{cl}}{d\ln a}=-(1+z)\frac{d\ln
  \delta_{cl}}{dz}\,,
\end{equation}
and 

\begin{equation}
\sigma_8(z)=\sigma_{80}\frac{\delta_{cl}(z)}{\delta_{cl}(z=0)}\,,
\end{equation} 
is the redshift-dependent root-mean-square mass fluctuation in spheres
with radius $8h^{-1}$ Mpc.

\begin{table}[ht]
  \begin{center}
    \begin{tabular}{l|l|l|l}
    \hline \;\;\ $z$ & $\;\;\; \;\; f\sigma_8$ &\;\;\; Survey &
    \;\;\;Ref.\\ \hline \hline 0.07 & $0.42 \pm 0.06$ & \;\; 6dFGRS &
    \;\; \cite{beutler}\\ 0.17 & $0.51 \pm 0.06$ & \;\; 2dFGRS & \;\;
    \cite{percival}\\ 0.22 & $ 0.42 \pm 0.07 $ & \;\; WiggleZ &\;\;
    \cite{blake}\\ 0.25 & $0.35 \pm 0.06$ & \;\; SDSS LRG &\;\;
    \cite{samushia2}\\ 0.37 & $0.46 \pm 0.04$ & \;\; SDSS LRG & \;\;
    \cite{samushia2}\\ 0.41 & $ 0.45 \pm 0.04 $ &\;\; WiggleZ &\;\;
    \cite{blake}\\ 0.57 & $0.43 \pm 0.03$ & \;\; BOSS CMASS
    &\;\;\ \cite{reid}\\ 0.60 & $ 0.43 \pm 0.04$ & \;\; WiggleZ & \;\;
    \cite{blake}\\ 0.78 & $ 0.38 \pm 0.04$ & \;\; WiggleZ &\;\;
    \cite{blake}\\ 0.80 & $0.47 \pm 0.08$ & \;\; VIPERS & \;\;
    \cite{delatorre}\\ \hline
      \end{tabular}
    \end{center}
    \caption{Observational data for redshift-space-distortion based
      $f(z)\sigma_8(z)$ and the sources from where we have obtained
      them.}
    \label{fsig8}
\end{table}

{}For the $f\sigma_8$ test we use the following $\chi^2$ statistics:

\begin{equation}
	\chi^{2}_{f\sigma_8} =
        \sum_{i=1}^{10}\frac{[f\sigma_8^{obs}(z_i)
            -f(z_i,\beta)\sigma_8(z_i,\sigma_{80},\beta)
          ]^2}{\sigma^2_{f\sigma_8}(z_i)}\,.
\end{equation}
To obtain the probability distributions (PDFs) in all the considered
tests in this work, the Metropolis-Hasting algorithm has been
used~\cite{gregory}. Generally, to obtain the PDFs, 40 chains were
generated with $10^6$ points for each chain. 

\begin{figure}[t]
  \begin{centering}
    \includegraphics[width=0.49\textwidth]{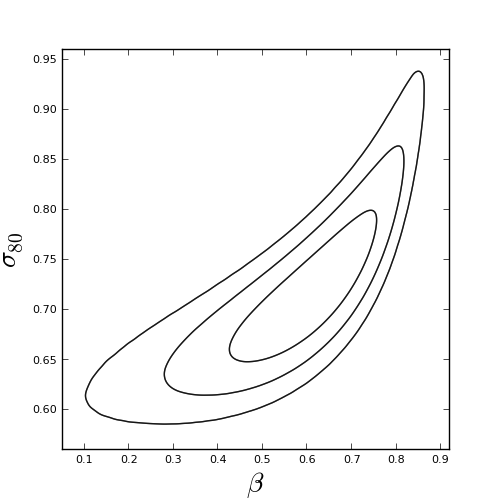}
    \includegraphics[width=0.49\textwidth]{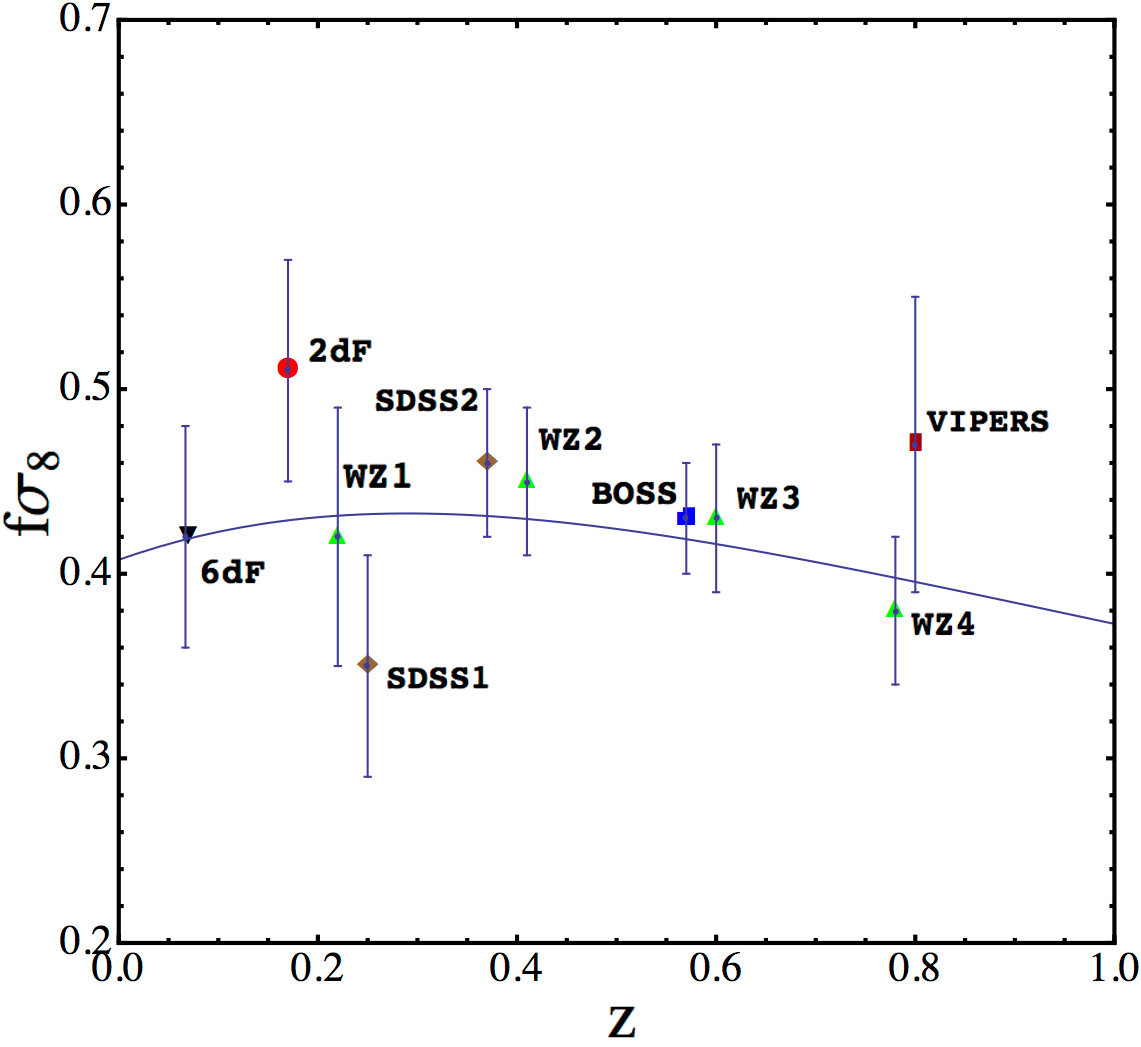}
    \par \end{centering}    \caption{Results for the $f\sigma_8$
    test. Left panel: Confidence regions in the $(\beta,\sigma_{80})$
    plane. {}From the outer to inner curves: Regions of 99.7$\%$,
    95.5$\%$ and 68.3$\%$C.L. Right panel: The $f\sigma_8$ data
    points (from Table~\ref{fsig8}) and the best fit model curve
    $f\sigma_8$, as a function of redshift.}
    \label{fig5}
\end{figure}

\begin{figure}[t]
\begin{centering}
    \includegraphics[width=1.1\textwidth]{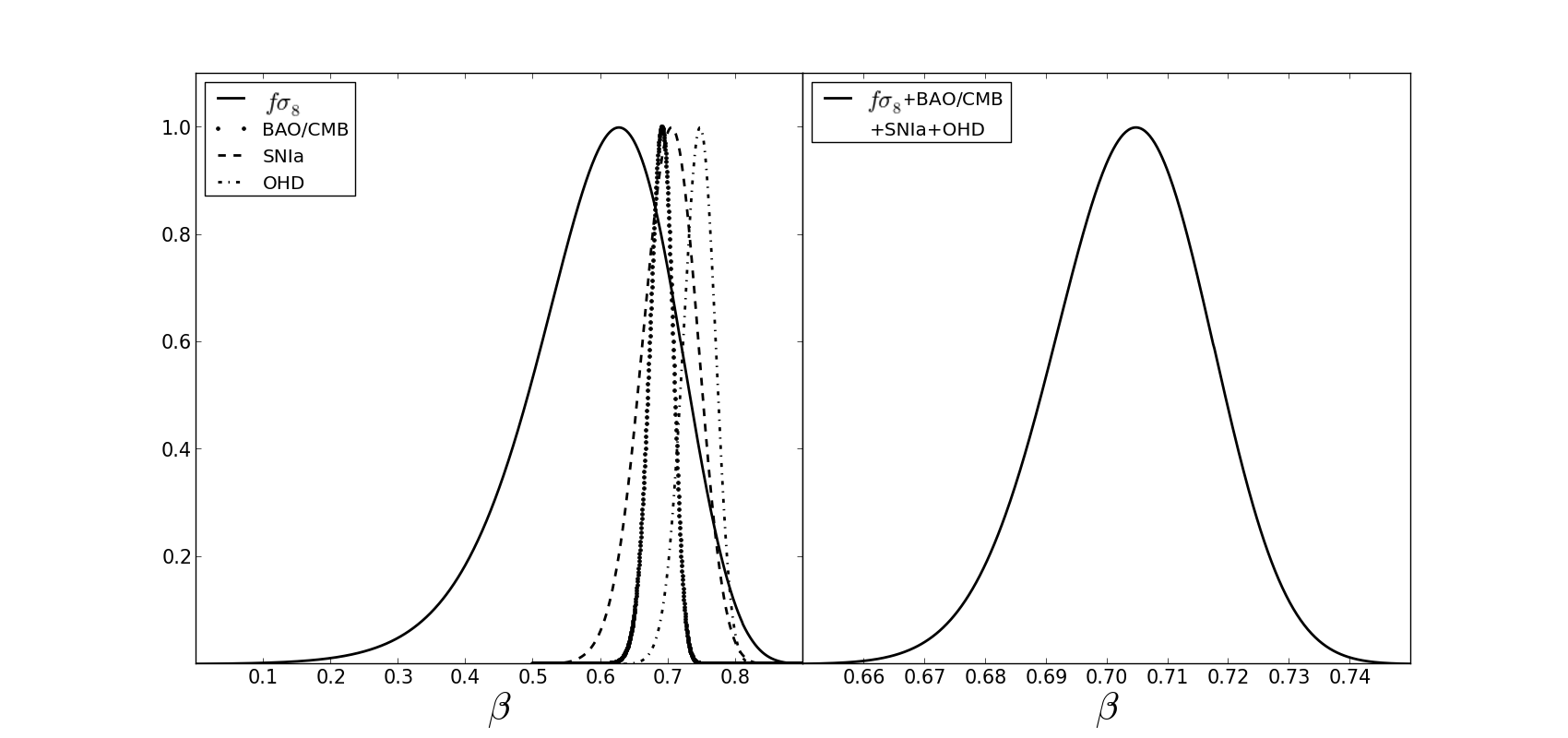} \par \end{centering}
    \caption{Left panel: The one-dimensional $\beta$ PDF for the
      redshift-space-distortion based $f(z)\sigma_8(z)$ test (solid
      curve) and for each background test, shown by the dotted, dashed and
      dash-dotted curves. Right panel: The one-dimensional PDF for $\beta$ 
      obtained by combining the previous four tests.}
    \label{fig6}
\end{figure}

The results for the $f\sigma_8$ test are displayed in
{}Fig.~\ref{fig5} (left panel). {}For this test, we obtain that  $
\beta = 0.63^{+ 0.09 (0.17)}_{-0.12 (0.26)} $, and
$\sigma_{80}=0.70^{+0.05(0.11)}_{-0.04(0.07)}$.  In {}Fig.~\ref{fig5}
(right panel) we also show the $f\sigma_8$ data points we have used,
along with their respective error bars, given by the results shown in
Table~\ref{fsig8}, and $f\sigma_8 $ for the best fit model, as a
function of redshift.  In {}Fig.~\ref{fig6}, after a flat
marginalization with respect to $\sigma_{80}$, we show the 
one-dimensional PDF for $\beta$ (given by the solid curve). 

{}For the background tests, which involve essentially only distances
and, thus, are independent of the perturbation results, we use the
following observables: (i) {\it The Union 2.1 Type Ia Supernovae
compilation}~\cite{suzuki} -- this compilation is an update of the
Union~2~\cite{amanullah} and include supernovae observed by the Hubble
Space Telescope Cluster Survey. This compilation is composed of 580
selected supernovae fitted by the SALT2-1 lightcurve fitter~\cite{site}.   
In our approach we have
considered the covariance matrix with systematics errors  (available
in the site mentioned in~\cite{site}), obtaining $\beta =
0.70^{+0.04(0.08)}_{-0.04(0.09)}$. (ii) {\it The CMB/BAO test} -- we followed
the procedure described in Sec. 3.2 of Ref.~\cite{giostri},
including one new data point from the BOSS survey~\cite{anderson} and
new data from WMAP-9yrs~\cite{wmap9}. With this test we get $\beta =
0.69^{+0.02(0.03)}_{-0.02(0.04)}$. (iii) {\it Measurements of the Hubble
parameter at different redshifts} -- for this observable we use the
same data set and procedure as described in Ref.~\cite{farook} and we
obtain $\beta = 0.75^{+0.02(0.04)}_{-0.02(0.05)}$. 

In the left panel of {}Fig.~\ref{fig6}, besides the result for the
$f\sigma_8$ test (solid curve), we also display the one-dimensional PDF
for $\beta$ for each background test: CMB/BAO (dotted curve),
SNeIa (dashed curve) and OHD (dash-dotted curve). 
We also display in {}Fig.~\ref{fig6}
(right panel) the $\beta$ one-dimensional PDF for the combined
$f\sigma_8$ plus the three background tests, which gives $\beta =
0.71^{+0.01(0.02)}_{ -0.01(0.03)}$.  

At this point it is important to make the following remark.  If we
have considered $c_{eff}^2\neq 0$, instead of Eq.~(\ref{delta_a}), we
would get from Eq.~(\ref{rdelta}) the following differential equation

\begin{equation}
\delta'' + \frac{3}{2a}\left(1-5w \right) \delta' + \frac{3}{2a^2}
\left( 3 w^2-8w -1 \right)\delta + \frac{c_{eff}^2k^2}{H^2a^4} \delta=
0. 
\label{delta_ak}
\end{equation}
It can be shown that below the Jeans length,
$\lambda_J=\sqrt{|c_{eff}^2|\pi/G\rho}$,  the $k$ dependence of the last
term in the left-hand side of the above equation can cause strong
oscillations if $c_{eff}^2 > 0$, or exponential growth if $c_{eff}^2 <
0$. Only models with $|c_{eff}^2|\ll 1$ are acceptable at linear
scales. An interesting question, but that is beyond the scope of this
paper, is to estimate upper limits that will  be imposed on $c_{eff}^2$
by future surveys like Euclid. 

In our approximation, we have not considered the presence of baryons. 
If we had taken them into account, still assuming $c_{eff}^2 =0$, it can be shown that their
density contrast has the same dependence with redshift as clustered dark matter (given 
by Eq.~(\ref{deltam})). Since the dependence with redshift of both energy densities 
is also the same, it will not be possible to distinguish the CCDM scenario from 
$\Lambda$CDM, by using measurements of the gas mass fraction in clusters \cite{mantz}, 
as suggested in Ref.~\cite{lima2}. 

\section{Conclusions}
\label{conclusions}

In this paper, we have studied the CCDM scenario as a possible
explanation for the late-time cosmic acceleration.  We have compared
the relativistic and neo-Newtonian differential equations for the
density  contrast for the CCDM model.  Both relativistic and
neo-Newtonian cases agree with each other only when the effective
sound speed $c_{eff}^2$ is equal to zero. We have argued that even in
the more general case,  in which $c_{eff}^2$ is considered
nonvanishing, but the momentum dependent term in the equation for the
density contrast is neglected, a somewhat common consideration assumed
by some authors, that this is also not a consistent approximation for
the density contrast differential equation.  This approximation of
neglecting the momentum dependent term is only justifiable  in the
long-wavelength limit ($k=0$), which is in turn exactly the case where the
Newtonian approximation is not valid. Thus, the neo-Newtonian approach
is not consistent with the full relativistic equations when $c_{eff}^2
\ne 0$, and the Newtonian approximation is not valid in the
long-wavelength limit.

Next, we have compared the CCDM predictions at the perturbative level
with those obtained from the $\Lambda CDM$. We have used for this
comparison  redshift-space-distortion observational data. We have
shown that the CCDM model produces results  for the  parameter $\beta$
(that at the background level plays the role of the cosmological
constant density parameter) that are fully consistent with the ones
expected from $\Lambda CDM$. Independent tests were also carried out
at the background level. These tests show that the result for $\beta$
predicted by the CCDM models is also consistent with  the result from
$\Lambda$CDM. We pointed out that this consistency with the
$\Lambda$CDM can only be achieved when we properly identify the
clustering part ($\delta_{cl}$) of the density contrast, as analyzed  and
argued in Sec.~\ref{ccdm}.  This subtle issue concerning the density
contrast may be related to the difficulties with the CCDM model found
in previous works.  {}For example, the authors of Ref.~\cite{roany}
have found that CCDM models tend to overestimate peculiar velocities
of galaxies in the linear regime. They have also found that because
the density contrast today, obtained from the $\Lambda$CDM model,
tends to be higher than the one predicted by the CCDM, that this would
result also in an overestimate of the present density of massive
galaxies clusters in these alternative models.  This result comes as a
consequence of the density contrast suppression as we increase $\beta$
and shown in the previous section. However, this difficult is no
longer present when the clustering part $\delta_{cl}$ is used instead.
As pointed out in Sec.~\ref{ccdm}, the
matter in CCDM clusters exactly in the same manner as it does  in
$\Lambda$CDM, since the gravitational potential is the same. 

In summary, we have shown that CCDM models with $c_{eff}^2=0$
are degenerate with $\Lambda$CDM not only
at the background level, but also at the linear perturbative level.
We can generally expect this degeneracy to remain at higher order.
Although $\Lambda$CDM has several conceptual problems  (smallness of
$\Lambda$, cosmic coincidence problem, etc), the CCDM model does not solve
any of them either. Therefore, in the absence of a more fundamental
microscopic basis for the particles creation rate and that originates
the specific CCDM model treated here, $\Lambda$CDM is a simpler alternative 
to explain observations (Occam's Razor).


\acknowledgments

 The authors would like to thank Julio Cesar Fabris, Jose Ademir Sales de 
Lima and Jose Antonio de Freitas Pacheco for useful discussions. 
 R.O.R and I.W. are partially supported by Conselho Nacional de
 Desenvolvimento Cient\'{\i}fico e Tecnol\'ogico (CNPq). M.V.S. thanks
 the Brazilian research agency CNPq for support.  R.O.R is also
 partially supported by a research grant from  Funda\c{c}\~ao Carlos
 Chagas Filho de Amparo \`a Pesquisa do Estado do Rio de Janeiro
 (FAPERJ).



\begin{thebibliography}{99}

\bibitem{DEreviews}   M.~Li, X.~-D.~Li, S.~Wang and Y.~Wang,
  Commun.\ Theor.\ Phys.\  {\bf 56}, 525 (2011);
 S.~Tsujikawa,
  arXiv:1004.1493 [astro-ph.CO].
  
\bibitem{Zeldovich:1970si} 
  Y.~.B.~Zeldovich,
  Pisma Zh.\ Eksp.\ Teor.\ Fiz.\  {\bf 12}, 443 (1970).

\bibitem{Murphy:1973zz} 
  G.~L.~Murphy,
  Phys.\ Rev.\ D {\bf 8}, 4231 (1973).

\bibitem{hu1982}B. L. Hu, 
Phys. Lett. A {\bf 90}, 375 (1982).

\bibitem{weinberg}S. Weinberg, {\it Gravitation and cosmology}
(Wiley, New York, 1972).

\bibitem{othersbulk}
  L.~Diosi, B.~Keszthelyi, B.~Lukacs and G.~Paal,
  Acta Phys.\ Polon.\ B {\bf 15}, 909 (1984);
  I.~Waga, R.~C.~Falcao and R.~Chanda,
  Phys.\ Rev.\ D {\bf 33}, 1839 (1986);
  J.~D.~Barrow,
  Nucl.\ Phys.\ B {\bf 310}, 743 (1988);
  R.~Maartens,
  Class.\ Quant.\ Grav.\  {\bf 12}, 1455 (1995);
   W.~Zimdahl, D.~Pavon and J.~Triginer,
  Helv.\ Phys.\ Acta {\bf 69}, 225 (1996);
  W.~Zimdahl and D.~Pavon,
  Phys.\ Rev.\ D {\bf 61},  108301 (2000).
   
 \bibitem{lima88}  J.~A.~S.~Lima, R.~Portugal and I.~Waga,
  Phys.\ Rev.\ D {\bf 37},  2755 (1988).

\bibitem{bulkDE} 
M.~Giovannini,
  Class.\ Quant.\ Grav.\  {\bf 22},  5243 (2005);
  S.~del Campo, R.~Herrera, D.~Pavon,
  Phys.\ Rev.\  D {\bf 75 },   083518  (2007)
C.~Bogdanos, A.~Dimitriadis and K.~Tamvakis,
Phys.\ Rev.\ D {\bf 75},  087303 (2007);
 B.~Li and J.~D.~Barrow,
  Phys.\ Rev.\ D {\bf 79},  103521 (2009).
J.~-S.~Gagnon and J.~Lesgourgues,
JCAP {\bf 1109},  026 (2011);
I.~Brevik, E.~Elizalde, S.~Nojiri and S.~D.~Odintsov,
Phys.\ Rev.\ D {\bf 84},  103508 (2011);
O.~F.~Piattella, J.~C.~Fabris and W.~Zimdahl,
JCAP {\bf 1105},  029 (2011).

\bibitem{prigogine}   E.~Gunzig, J.~Geheniau and I.~Prigogine,
  Nature {\bf 330},  621 (1987) ;
I. Prigogine, J. Geheniau, E. Gunzig and P. Nardone, 
Proc. Natl. Acad. Sci. USA {\bf 85},  7428 (1988);
  I.~Prigogine, J.~Geheniau, E.~Gunzig and P.~Nardone,
  Gen.\ Rel.\ Grav.\  {\bf 21},  767 (1989).
    
\bibitem{calvao}  M.~O.~Calvao, J.~A.~S.~Lima and I.~Waga,
  Phys.\ Lett.\ A {\bf 162},  223 (1992).
    
  \bibitem{pacher}  T.~Pacher, J.~A.~Stein-Schabes and M.~S.~Turner,
  Phys.\ Rev.\ D {\bf 36},  1603 (1987).

\bibitem{mattercreation}   V.~H.~Cardenas,
  Eur.\ Phys.\ J.\ C {\bf 72},  2149 (2012);
  S.~K.~Modak and D.~Singleton,
  Phys.\ Rev.\ D {\bf 86}, 123515 (2012);
  S.~Debnath and A.~K.~Sanyal,
  Class.\ Quant.\ Grav.\  {\bf 28}, 145015 (2011);
  G.~Steigman, R.~C.~Santos and J.~A.~S.~Lima,
  JCAP {\bf 0906}, 033 (2009).

\bibitem{freaza}  M.~P.~Freaza, R.~S.~de Souza and I.~Waga,
  Phys.\ Rev.\ D {\bf 66},  103502 (2002).

\bibitem{zimdahl}  W.~Zimdahl, D.~J.~Schwarz, A.~B.~Balakin and D.~Pavon,
  Phys.\ Rev.\ D {\bf 64},  063501 (2001).

\bibitem{lima1}   J.~A.~S.~Lima, J.~F.~Jesus and F.~A.~Oliveira,
  JCAP {\bf 11},  027 (2010).

\bibitem{basilakos}  S.~Basilakos and J.~A.~S.~Lima,
  Phys.\ Rev.\ D {\bf 82},  023504 (2010).

\bibitem{jesus}  J.~F.~Jesus, F.~A.~Oliveira, S.~Basilakos and J.~A.~S.~Lima,
  Phys.\ Rev.\ D {\bf 84},  063511 (2011).

\bibitem{lima2}J.~A.~S.~Lima, S.~Basilakos and F.~E.~M.~Costa,
  Phys.\ Rev.\ D {\bf 86},  103534 (2012).

\bibitem{BMR}
  A.~Berera, I.~G.~Moss and R.~O.~Ramos,
  Rept.\ Prog.\ Phys.\  {\bf 72},  026901 (2009).

\bibitem{warmbulk}  M.~Bastero-Gil, A.~Berera, R.~Cerezo, R.~O.~Ramos and G.~S.~Vicente,
  JCAP {\bf 1211},  042 (2012).

\bibitem{lima3}  J.~A.~S.~Lima, V.~Zanchin and R.~H.~Brandenberger,
  Mon.\ Not.\ Roy.\ Astron.\ Soc.\  {\bf 291},  L1 (1997)

\bibitem{reis}  R.~R.~R.~Reis,
  Phys.\ Rev.\ D {\bf 67},  087301 (2003)
   [Erratum-ibid.\ D {\bf 68},  089901 (2003)].

\bibitem{kodama}  H.~Kodama and M.~Sasaki,
  Prog.\ Theor.\ Phys.\ Suppl.\  {\bf 78},  1 (1984).

\bibitem{milne}E. A. Milne, Quarterly J. Math. \textbf{5},  64 (1934); 
republished in Gen. Rel. Gravit. \textbf{32},  1939 (2000).

\bibitem{mccrea}W. H. McCrea and E. A. Milne, Quarterly J. Math. \textbf{5},  73 (1934); 
republished in Gen. Rel. Gravit. \textbf{32},  1949 (2000).

\bibitem{mccrea2} W. H. McCrea, Proc. R. Soc. London \textbf{206},  562 (1951).

\bibitem{harrison2} E. R. Harrison, Ann. Phys. (N.Y.) \textbf{35},  437 (1965).

\bibitem{sachs}  R.~K.~Sachs and A.~M.~Wolfe,
  Astrophys.\ J.\  {\bf 147},  73 (1967)
   [Gen.\ Rel.\ Grav.\  {\bf 39},  1929 (2007)].

\bibitem{hu2} 
 W. Hu, Summer School Lectures: 2002 Astroparticle Physics and Cosmology, eds. G. Dvali et al. (Abdus Salam ICTP, Trieste, 2003) p. 149 
 [arXiv:0402060 [astro-ph.CO]]; W. Hu,  Astrophys.\ J.\  {\bf 506},  485 (1998).

\bibitem{silveira} V. Silveira and I. Waga, Phys. Rev. D {\bf{50}},  4890 (1994).

\bibitem{roany}  A.~de Roany and J.~A.~d.~F.~Pacheco,
  Gen.\ Rel.\ Grav.\  {\bf 43},  61 (2011).
  
\bibitem{kunz} M. Kunz,  Phys.\ Rev.\ D {\bf 80},  123001 (2009).
 
\bibitem{fabris} J. C. Fabris, S. V. B. Goncalves and R. S. Ribeiro,  Gen.\ Rel.\ Grav.\  {\bf 36}, 211 (2004).
 
 \bibitem{avelino} P. P. Avelino, L. M. G. Beca, J. P. M. de Carvalho and C. J. A. P. Martins, 
JCAP {\bf 09},  002 (2003).
 
\bibitem{song}  Y.~-S.~Song and W.~J.~Percival,
  JCAP {\bf 0910},  004 (2009).

 \bibitem{beutler}  F.~Beutler, {\it et al.},
Mon. Not. R. Astron. Soc. {\bf 423}, 3430 (2012).
 
\bibitem{percival}  W.~J.~Percival, {\it et al.}  [2dFGRS Collaboration],
  Mon.\ Not.\ Roy.\ Astron.\ Soc.\  {\bf 353},  1201 (2004).
  
\bibitem{blake}  C.~Blake, {\it et al.},
  Mon.\ Not.\ Roy.\ Astron.\ Soc.\  {\bf 415},  2876 (2011).

\bibitem{samushia2}  L.~Samushia, W.~J.~Percival and A.~Raccanelli,
  Mon.\ Not.\ Roy.\ Astron.\ Soc.\  {\bf 420},  2102 (2012).
 
\bibitem{reid} B. A. Reid \textit{et al.},  
  Mon.\ Not.\ Roy.\ Astron.\ Soc.\  {\bf 426},  2719 (2012).

\bibitem{delatorre}  S.~de la Torre, {\it et al.},
Astron. Astrophys. {\bf 557}, A54 (2013).
 
\bibitem{gregory} P. C. Gregory, {\it Bayesian Logical Data Analysis for the Physical Sciences: 
a Comparative Approach with Mathematica Support} 
(Cambridge University Press, Cambridge, UK, 2005); 
D. Gamerman and H. F. Lopes, {\it Markov Chain Monte Carlo: Stochastic Simulation for 
Bayesian Inference} (Chapman \& Hall/CRC, 2006).
  
\bibitem{suzuki}  N.~Suzuki, {\it et al.},
  Astrophys.\ J.\  {\bf 746},  85 (2012).

\bibitem{amanullah}  R.~Amanullah,  {\it et al.},
  Astrophys.\ J.\  {\bf 716},  712 (2010).

\bibitem{site}http://supernova.lbl.gov/Union

\bibitem{giostri}  R.~Giostri,  {\it et al.},
  JCAP {\bf 1203},  027 (2012).

\bibitem{anderson}  L.~Anderson,  {\it et al.},
  Mon.\ Not.\ Roy.\ Astron.\ Soc.\  {\bf 427},  3435 (2013).

\bibitem{wmap9} http://lambda.gsfc.nasa.gov/product/map/current/parameters.cfm

\bibitem{farook}  O.~Farooq and B.~Ratra,
  Astrophys.\ J.\  {\bf 766},  L7 (2013).
  
 \bibitem{mantz}A. B. Mantz, S. W. Allen, R. G. Morris, D. A. Rapetti,
D. E. Applegate, P. L. Kelly, A. von der Linden and R. W. Schmidt, arXiv:1402.6212 [astro-ph.CO].

\end{thebibliography}
\end{document}